\documentclass[reprint,amsmath,amssymb,aps,prb,twocolumn,superscriptaddress,showpacs]{revtex4}

\makeatletter
\usepackage{graphicx}
\usepackage{dcolumn}
\usepackage{bm}
\usepackage{amsmath,graphicx,latexsym}
\usepackage{setspace}
\usepackage{multirow}
\usepackage{verbatim}   
\usepackage{array}

\def\LCO{{LaCrO$_3$}}
\def\LFO{{LaFeO$_3$}}
\def\SMO{{SrMnO$_3$}}
\def\LCFO{{La$_2$CrFeO$_6$}}
\def\CMO{{CaMnO$_3$}}
\def\BMO{{BaMnO$_3$}}
\def\G4{{$\Gamma_4^-$}}
\def\R5{{R$_5^-$}}
\def\cm1{{cm$^{-1}$}}
\def\GGAU{{GGA+$U$}}
\def\AMO{{$A$MnO$_3$}}
\def\LMO{{La$M$O$_3$}}

\begin{document}

\title{Spin-phonon coupling effects in transition-metal perovskites:\\
a DFT+$U$ and hybrid-functional study}

\author{Jiawang Hong}
\email{hongjw10@physics.rutgers.edu}
\affiliation{ Department of Physics and Astronomy, Rutgers University,
 Piscataway, NJ 08854-8019, USA }

 \author{Alessandro Stroppa}
 \affiliation{CNR-SPIN, L'Aquila, Italy}

 \author{Jorge \'I\~niguez} 
 \affiliation{Institut de Ci\`encia de Materials de
 Barcelona (ICMAB-CSIC), Campus UAB, 08193 Bellaterra, Spain }

 \author{Silvia Picozzi}
 \affiliation{CNR-SPIN, L'Aquila, Italy}

\author{David Vanderbilt}
\affiliation{ Department of Physics and Astronomy, Rutgers University,
 Piscataway, NJ 08854-8019, USA }

\date{\today}

\begin{abstract}
Spin-phonon coupling effects, as reflected in phonon frequency shifts
between ferromagnetic (FM) and G-type antiferromagnetic (AFM)
configurations in cubic CaMnO$_3$, SrMnO$_3$, BaMnO$_3$, LaCrO$_3$,
LaFeO$_3$ and La$_2$(CrFe)O$_6$, are investigated using
density-functional methods.  The calculations are carried out both
with a hybrid-functional (HSE) approach and with a DFT+$U$ approach
using a $U$ that has been fitted to HSE calculations.  The phonon
frequency shifts obtained in going from the FM to the AFM spin
configuration agree well with those computed directly from the more
accurate HSE approach, but are obtained with much less computational
effort.  We find that in the $A$MnO$_3$ materials class with $A$=Ca,
Sr, and Ba, this frequency shift decreases as the A cation radius
increases for the $\Gamma$ phonons, while it increases for R-point
phonons.  In La$M$O$_3$ with $M$=Cr, Fe, and Cr/Fe, the phonon
frequencies at $\Gamma$ decrease as the spin order changes from AFM to
FM for LaCrO$_3$ and LaFeO$_3$, but they increase for the double
perovskite La$_2$(CrFe)O$_6$.  We discuss these results and the
prospects for bulk and superlattice forms of these materials to be
useful as multiferroics.
\end{abstract}


\maketitle


\section{Introduction}
\label{sec:intro}

Multiferroic materials are compounds showing coexistence of two or
more ferroic orders, e.g.,
ferroelectricity together with some form of magnetic
order such as ferromagnetism or antiferromagnetism.
Presently such materials are attracting enormous
attention due to their potential for advanced device
applications and because they offer a rich playground
from a fundamental physics point of view. Possible applications
tend to focus on the magnetoelectric (ME) coupling, which may pave
the way for control of  the magnetization by an applied electric
field in spintronic devices,\cite{Appl2,Appl3} although other
applications such as four-state memories \cite{Appl1} are also
of interest.

Although intrinsic multiferroic materials are highly desirable, they
are generally scarce.  One often-proposed reason may be that partially
filled 3$d$ shells favor magnetism, while the best-known ferroelectric
(FE) perovskites have a 3$d^{0}$ configuration for the transition
metal (e.g.  Ti$^{4+}$, Nb$^{5+}$, etc.\cite{Few}). However, it has
recently been shown \cite{Ghosez2} that some magnetic perovskite
oxides display incipient or actual FE instabilities, clearly
indicating that there are ways around the usually-invoked
incompatibility between ferroelectricity and magnetism. Some of these compounds,
namely \CMO, \SMO\ and \BMO, will be considered in this work.

Even if several microscopic mechanisms have been identified for the
occurrence of ferroelectricity in magnetic materials, \cite{KM,Ederer}
alternative routes are needed in order to optimize materials for
functional device application.  This could be done either by focusing
on new classes of compounds such as hybrid organic-inorganic
materials\cite{Routes1, Routes2,AngewStroppa} or by modifying
already-known materials to engineer specific properties.  In the
latter direction, an intriguing possibility is to start with non-polar
materials and then induce multiple non-polar instabilities; under
appropriate circumstances this can produce a ferroelectric
polarization, as first predicted in Ref.~\onlinecite{Ref29} based on
general group theory arguments and analyzed in the
SrBi$_{2}$Nb$_{2}$O$_{9}$ compound by means of a symmetry analysis
combined with density-functional theory calculations by Perez-Mato
\textit{et al.}\cite{Ref30} Here the ferroelectricity was found to
arise from the interplay of several degrees of freedom, not all of
them associated with unstable or nearly-unstable modes. In particular,
a coupling between polarization and two octahedral-rotation modes was
invoked to explain the behavior.\cite{Ref30} Bousquet \textit{et al.}
have demonstrated that ferroelectricity is produced by local
rotational modes in a SrTiO$_{3}$/PbTiO$_{3}$
superlattice.\cite{Ref31}
Although in most improper ferroelectrics a single primary
order parameter induces the polarization,\cite{Ref32}
Benedek and Fennie proposed that the
combination of two lattice rotations, neither of which produces
ferroelectric properties individually, can induce a ME coupling, weak
ferromagnetism, and ferroelectricity.\cite{Ref33}
Indeed, we now know
that rotations of the oxygen octahedra, in
combination\cite{Ref33,Ref34,Ref35} and even
individually,\cite{ederer06,Jorge1} can produce ferroelectricity,
modify the magnetic order, and favor magnetoelectricity.

Another route to creating new multiferroic materials may be to exploit
the coupling between polarization, strain, and spin degrees of
freedom.  A strong dependence of the lowest-frequency polar phonon
frequency on epitaxial strain\cite{Cohen} is common in paraelectric
(PE) perovskite oxides, and can sometimes be exploited to drive the
system ferroelectric, a phenomenon known as epitaxial-strain--induced
ferroelectricity.\cite{StrainSTO} In a magnetic system that also has a strong
spin-phonon coupling, i.e., a strong dependence of the polar
phonon frequencies on spin order, the magnetic order may be capable of
tipping the balance between PE and FE states.  For example, consider a
system that has two competing ground states, one of which is
antiferromagnetic (AFM) and PE while the other is ferromagnetic (FM)
and FE, and assume that the spin-phonon coupling is such that the
lowest-frequency polar phonon is softer for FM ordering than for AFM
ordering.  In such a case, the epitaxial-strain enhancement of a polar
instability may lead to a lowering of the energy of the FM-FE state
below that of the AFM-PE phase. Spin-phonon mechanisms of this kind
have been powerfully exploited for the design of novel multiferroics
in Refs.~\onlinecite{Craig1,Nature1,
strain1,strain2,strain3,strain4, Mochizuki}.

Clearly,  it is desirable to have a large spin-phonon
coupling, in terms of a large shift $\Delta \omega$ of phonon
frequencies upon change of the magnetic order.
Interestingly, recently computed spin-phonon couplings in
ME compounds seems to be strikingly large. For
example, a $\Delta \omega$ of about 60 cm$^{-1}$ has been
reported for the double perovskite
La$_{2}$NiMnO$_{6}$.\cite{Das1} Furthermore, $\Delta \omega$
values of about 200 cm$^{-1}$ have been computed for a cubic phase of
SrMnO$_{3}$.\cite{Junhee1}  These values appear  to be
anomalously large when compared to phonon splittings across
magnetic phase transitions in other oxides, which are in the 5-30
cm$^{-1}$ range.~\cite{Hemberg,Rudolf}

In the search for materials with large spin-phonon
coupling, first-principles based calculations play a prominent
role since they can pinpoint promising candidates
simply by inspecting the dependence of the
phonon frequencies on the magnetic order.  However, a serious
bottleneck appears. To obtain a reliable and accurate
description of these effects,
it is important to describe the structural,
electronic and magnetic properties on an equal footing. This is
especially true for transition-metal oxides, which are the usual
target materials for the spin-phonon driven
ferroelectric-ferromagnet. In this case, it is well known that
the localized nature of the 3$d$ electronic states, or loosely
speaking, the ``correlated'' nature of these compounds,  poses
serious limits to the applicability of common density-functional
methods like the local density approximation (LDA) or generalized
gradient approximation (GGA). In fact, these standard approximations
introduce a spurious Coulomb interaction of the electron
with its own charge, i.e.,
the electrostatic self-interaction is not entirely compensated.
This causes fairly large errors for localized states (e.g.,
Mn \textit{d} states).  It tends to destabilize the orbitals and
decreases their binding energy, leading to an overdelocalization
of the charge density.\cite{Kronik}

One common way out is the use of the DFT+$U$ method, \cite{GGA+U}
where a Hubbard-like $U$ term is introduced into the DFT energy
functional in order to take correlations partially into account.  The
method usually improves the electronic-structure description, but it
suffers from shortcomings associated with the $U$-dependence of the
calculated properties.\cite{KresseDFTU} Unfortunately, there is
usually no obvious choice of the $U$ value to be adopted; common
choices are usually based either on experimental input or are derived
from constrained DFT calculations, but neither of these approaches is
entirely satisfactory.  When dealing with phonon calculations, the
$U$-dependence becomes even more critical since the phonon frequencies
depend strongly on the unit-cell volume used for their evaluation, and
the theoretical volume, in turn, depends on $U$.  It is worth
mentioning that even if an appropriate choice of $U$ can accurately
reproduce the binding energy of localized $d$ states of
transition-metal oxides, it is by no means guaranteed that the same
$U$ can accurately reproduce other properties of the same compound,
such as the volume.\cite{KresseDFTU,PCCP.BFO} While most papers
address the spin-phonon coupling by applying DFT+$U$ methods, only a
few deal with the dependence upon the $U$ parameter.\cite{Ray} In this
paper, we will show that the spin-phonon coupling can strongly depend
on the $U$ parameter, and that such a dependence may give rise to
artificially large couplings. It is important to stress this message,
often overlooked in the literature, in view of the increasing interest
in ab-initio predictions of ferroelectric materials driven by
spin-phonon coupling.

In the last few years, another paradigmatic approach has been
widely applied in solid-state materials science, namely the
use of ``hybrid functionals'' that incorporate a weighted mixture of
the exchange defined in the
Hartree-Fock theory (but using the Kohn-Sham orbitals) and the
density-functional exchange. The correlation term is retained
from the density-functional framework. 
It is now widely accepted that the hybrid functionals outperform
semilocal functionals, especially for bulk materials with band
gaps.\cite{hyb1,hyb2,hyb3,hyb4,hyb5,hyb6,hyb7,hyb8,hyb9,hyb10,hyb11,hyb12}
It has also been shown that for low-dimensional systems such
as semiconductor/oxides interfaces, the performance of hybrid
functionals remains quite satisfactory. \cite{Shaltaf} However,
some doubts have very recently been put forward
about the performance of hybrid functionals for certain structural
configurations, e.g., surfaces or nanostructures.\cite{Jain}

While there is  a plethora of
different hybrid functionals, many of them defined empirically,
a suitable functional derived on theoretical grounds
is the so called PBE0,\cite{PBE01,PBE02} where
 the exchange mixing parameter as been fixed
to one quarter as justified by a perturbation-theory calculation.
A closely related functional, the Heyd-Scuseria-Ernzerhof
(HSE) hybrid functional,\cite{HSE06}
 introduces yet another parameter $\mu$ which splits the Coulomb
interaction kernel into short- and long-range pieces while retaining
the mixing
only on the short-range component. It has been shown that
this new hybrid functional,~\cite{HSE06} while preserving most of the
improved performance of PBE0 with respect to standard
local and semi-local exchange correlation functionals, greatly reduces
the computational cost.  For this reason, it is especially suitable
for periodically extended systems,
and is currently being applied in many solid-state
applications ranging from simple semiconductor systems to
transition metals, lanthanides, actinides, molecules at surfaces,
diluted magnetic semiconductors, and carbon nanostructures (for a
recent review see Ref.~\onlinecite{HFrev}).  The HSE functional has
been also used for  phonon calculations for simple semiconducting
systems\cite{hseph1,hseph2} or perovskite
structures,\cite{hseph3,hseph4} where it was shown the that phonon modes
are much more accurately reproduced using the hybrid functionals than using
GGA or LDA.~\cite{Wahl} 

Very recently, extended benchmarkings of the HSE method as well as
other self-interaction--corrected approaches have been presented for
prototypical transition metal oxides such as MnO, NiO and LaMnO$_3$,
and it has been shown that ``HSE shows a remarkable quantitative
agreement with experiments on most examined properties''\cite{Cesare}
and ``HSE shows predictive power in describing exchange interactions
in transition metal oxides.''\cite{Wahl2} These recent studies further
motivate us to use the HSE functional for our studies of the
spin-phonon coupling; indeed, as we will see in Sec.~\ref{subsec:CSB},
the HSE-calculated phonon modes agree well with available experimental
results for \SMO. The spin-phonon coupling effect, to the best of our
knowledge, is totally unexplored by hybrid functional approaches. In
this work we aim at filling this gap.

Even when using the more efficient HSE functional, however, the
use of hybrids entails an increased computational cost which
makes the calculation of phonon properties of complex magnetic
oxides very difficult. In this paper, we propose to circumvent this
bottleneck by combining the  HSE and DFT+$U$ approaches, i.e., choosing
the appropriate $U$ for each material by fitting to the HSE results
for some appropriate materials properties.  This provides
a fairly efficient and affordable strategy that preserves
the ``HSE accuracy''  for lattice constants, spin-phonon
couplings, and related properties, while taking advantage of the
computationally inexpensive DFT+$U$ method for the detailed
calculations.  Further details will be given in Section~\ref{sec:method}.

As far as the materials are concerned, it has been suggested that
\AMO\ perovskites with $A$\,=\,Ca, Sr and Ba may be good candidates
for spin-phonon--coupling driven
multiferroicity.\cite{Ghosez2,Junhee1,James} Furthermore, it has been
reported that both \SMO\ and \CMO\ have a large spin-phonon
coupling.\cite{Junhee1,Barone} We will show below that the spin-phonon
coupling can depend strongly on the chosen $U$.  We will also consider
the simple perovskite \LMO\ materials with $M$\,=\,Cr and Fe, which
have N\'eel temperatures above room temperature~\cite{Goodenough} and
large band gaps.  We want to explore the possibility of using these
two materials as building blocks for room temperature multiferroics,
e.g., in the form of double perovskites such as \LCFO.  We have chosen
these two classes of materials in part for the reasons outlined above,
but also because they are sufficiently ``easy-to-calculate'' for the
benchmark and testing purposes of the present work, especially when
considering hybrid functionals.

The paper is organized as follows. In Sec.~\ref{sec:method} we report
the computational details and describe our strategy for fitting $U$ of
the \GGAU\ calculations via a preliminary ``exploratory'' study using the
HSE method.  In Sec.~\ref{sec:results} we discuss: i) the  
effect of $U$ on the frequency shift, in Sec.~\ref{sec:U};
ii) the spin-phonon coupling effects
in $A$MnO$_3$ with $A$\,=\,Ca, Sr, and Ba, in Sec.~\ref{subsec:CSB};
iii) the spin-phonon coupling effects in La$M$O$_3$ with
$M$\,=\,Cr, Fe, and the corresponding double
perovskite La(Cr,Fe)O$_{3}$ in Sec.~\ref{sect:double}; and
iv) prospects for the design of new multiferroics in
Sec.~\ref{sect:multiferroics}.  In Sec.~\ref{sec:summ}, we
give a summary and conclusions. Finally, in the Appendix, we give details 
about the methodology used here for a full HSE phonon calculation.

\section{Methods}
\label{sec:method}

\begin{figure}[htpb]
  \begin{center}
   \includegraphics[width= 3.2in] {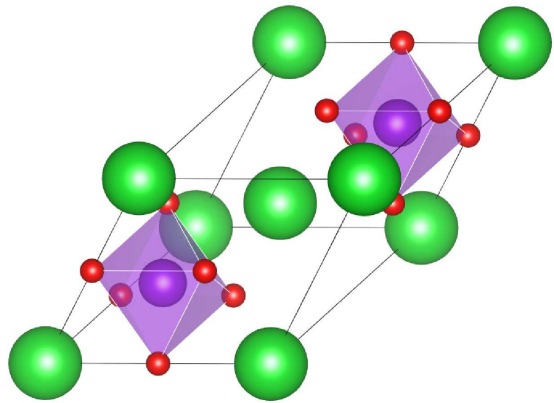}
  \end{center}
  \caption{(Color online) $AB$O$_3$ perovskite structure doubled along
    the [111] direction. $A$ atoms (largest) shown in green;
    $B$ atoms shown in violet; O atoms (smallest) shown in red.}
  \label{fig:structure}
\end{figure}

All of our calculations were performed in the framework of density-functional
theory as implemented in the Vienna \textit{ab-initio} simulation
package (VASP-5.2)~\cite{VASP1,VASP2} with a plane-wave cutoff of
500\,eV. The  DFT+$U$ and HSE calculations were carried out using the
same set of projector-augmented-wave (PAW) potentials 
to describe the electron-ion interaction.\cite{PAW1,PAW2} 
The unit cell for simulating the G-AFM magnetic ordering,
where all first-neighbor spins are antiferromagnetically aligned, is
doubled along the [111] direction as shown in
Fig.~\ref{fig:structure}.  The same simulation cell is retained
for the FM configuration in order to avoid numerical artefacts
that could arise in comparing calculations with different
effective k-point samplings. A $6\times 6 \times 6$ $\Gamma$-centered
$k$-point mesh is used.

\begin{figure*}[htpb]
  \begin{center}
 \includegraphics[width= 7.2 in] {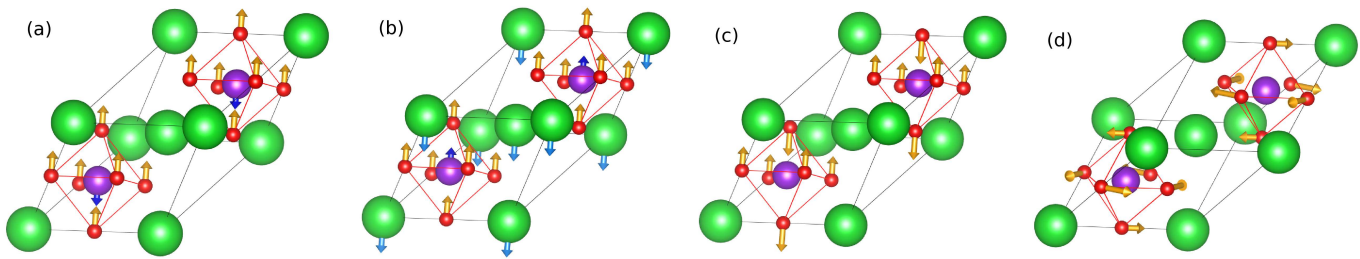}
  \end{center}
  \caption{(Color online) Three idealized polar modes and $R_5^-$
  mode in simple perovskites. (a) Slater mode; (b)
  Last mode; (c) Axe mode; (d) $R_5^-$ mode.}
  \label{fig:modes}
\end{figure*}

The phonon frequencies were calculated using the frozen-phonon method.
Except for the case of \LCFO, the charge density and dynamical matrix
retain the primitive 5-atom-cell periodicity for both the G-AFM and FM
spin structures, so it is appropriate to analyze the phonons using
symmetry labels from the primitive $Pm\bar 3m$ perovskite cell.  The
zone-center phonons decompose as $3\,\Gamma_4^-\oplus\Gamma_5^-$ (plus
acoustic modes), with the \G4\ modes being polar.  The zone-boundary
modes at the R point (which appear at the $\Gamma$ point in our
10-atom-cell calculations) decompose as ${\rm R}_5^+\oplus{\rm
  R}_5^-\oplus2\,{\rm R}_4^-\oplus{\rm R}_2^-\oplus{\rm R}_3^-$, where
the three-fold degenerate R$_5^-$ is of special interest because it
corresponds to rigid rotations and tilts of the oxygen octahedra.
After each frozen-phonon calculation of the zone-center phonons of our
10-atom cell, we analyze the modes to assign them according to these
$k$-point and symmetry labels.  For \LCFO, some of the modes mix
(e.g., \G4\ with R$_5^+$); in these cases we report the dominant mode
character.

As for the polar \G4\ phonons, these are often characterized
in $AB$O$_3$ perovskites
in terms of three kinds of idealized modes
as illustrated in Figs.~\ref{fig:modes}a-c.
The Slater mode (S-\G4) of Fig.~\ref{fig:modes}(a)
describes the vibration of $B$ cations against the oxygen octahedra;
the Last (L-\G4) mode of Fig.~\ref{fig:modes}(b) expresses a vibration
of the $A$ cations against the $B$O$_6$ octahedra; and the
Axe mode (A-\G4) of Fig.~\ref{fig:modes}(c) represents the distortion
of the oxygen octahedra. \cite{Hlinka}
The actual mode eigenvectors never
behave precisely like these idealized cases, but we find
that they can be identified in practice as being mainly of one
character, which is the one we report.
These polar modes contribute to the low-frequency dielectric constant,
and their softening in the high-symmetry paraelectric phase indicates
the presence of a ferroelectric instability.
For the insulating compounds, we further calculated their dielectric
constants and Born effective charges 
using density-functional perturbation theory within
the DFT+$U$ context as implemented in VASP.
The antiferrodistortive (AFD) mode ($R_5^-$ mode), which describes
the rotation oxygen octahedra,
is also shown in Fig.~\ref{fig:modes}(d).

We make use of the DFT+$U$ method~\cite{GGA+U} in the standard
Dudarev implementation where the on-site Coulomb interaction for
the localized $3d$ orbitals is parametrized by $U_{\rm eff}=U-J$
(which we denote henceforth as simply $U$)~\cite{Dudarev} using
the PBEsol functional,\cite{PBEsol} which has
been shown to give a satisfactory description of solid-state
equilibrium properties. We shall refer to this as PBEsol+{\sl U}.
The lack of experimentally available data for our systems
prevents us from extracting $U$ directly from experiments;
we will return to this delicate point shortly.

The other functional we have used is
HSE06,\cite{HSE06} a screened hybrid functional introduced by
Heyd, Scuseria, and Ernzerhof (HSE), where  one quarter of the
PBE short-range exchange is replaced by exact exchange, while
the full PBE correlation energy is included.  The
range-separation parameter $\mu$ is set to $\mu$ = 0.207\,\AA$^{-1}$.
The splitting of the Coulomb interaction into
short- and long-range pieces, as done in HSE, allows for a faster
numerical convergence with $k$-points when dealing with solid-state
systems.  However, as previously mentioned, the application
of the HSE approach to phonon calculations for magnetic oxide systems
remains very challenging in terms of computational workload.

\subsection*{Scheme to fit $U$ from hybrid calculations}

Here, we propose a practical scheme to perform relatively inexpensive
DFT+$U$ simulations that retain an accuracy comparable to HSE for the
calculation of spin-phonon couplings.

Our goal is to obtain a DFT+$U$ approach that reproduces the
dependence of the materials properties on the spin arrangement as
obtained from HSE calculations. Naturally, the first and most basic
property we would like to capture correctly is the energy itself. As
it turns out, the energy is also a very sensible property on which one
can base a $U$-fitting scheme: The energy differences between spin
configurations are directly related to the magnetic interactions or
exchange constants, and these are known to depend on the on-site
Coulomb repulsion $U$ affecting the electrons of the magnetic
species. (Typically, in our compounds of interest, the value of $U$
used in the simulations will play an important role in determining the
character of the top valence states; in turn, this will have a direct
impact on the nature and magnitude of the exchange couplings between
spins.)  This dependence of the exchange constants on $U$ makes this
criterion a very convenient one for our purposes. Of course, such a
fitting procedure does not guarantee our DFT+$U$ scheme will reproduce
correctly the phonon frequencies and frequency shifts between
different spin arrangements obtained from HSE calculations. In that
sense, we are relying on the physical soundness of the Hubbard-$U$
correction to DFT; as we will see below, the results are quite
convincing.

\begin{figure}[htpb]
  \begin{center}
    \includegraphics{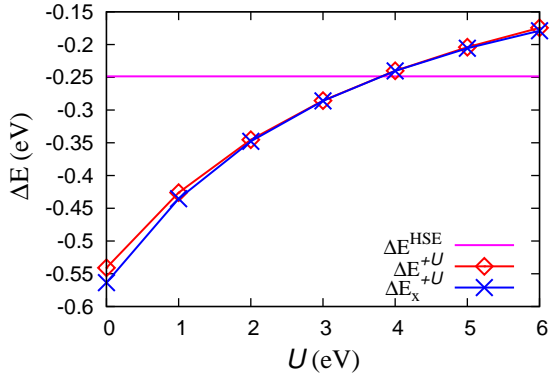}
  \end{center}
  \caption{(Color online) Variation of PBEsol+$U$ total energy
  difference $\Delta E=E_{\rm AFM}-E_{\rm FM}$ with $U$
  for \LCO, $\Delta E^{+U}$ is the energy difference with
  volume fixed to the optimized volume from HSE and
  $\Delta E_x^{+U}$ relaxed volume for each $U$. (The fitted value of
  $U$ occurs at the crossing with the $\Delta E^{\rm HSE}$ value.)}
  \label{fig:fitU}
\end{figure}

Obtaining $U$ from the energy differences has an additional advantage:
It allows us to devise a very simple fitting procedure that relies on
a minimal number of HSE calculations. In essence, these are the steps
we follow. We relax the cubic structure separately for G-AFM and FM
spin configurations using HSE and obtain the total-energy difference
$\Delta E$(HSE) =$E_{\rm AFM}$ (HSE) - $E_{\rm FM}$(HSE). We then
carry out a series of PBEsol+{\sl U} calculations in which $U$ is
varied from 0 to 6\,eV and obtain $\Delta E(+U)$ = $E_{\rm
  AFM}(+U)-E_{\rm FM}(+U)$ for each $U$. The $U$ is then chose such
that $\Delta E(+U)$= $\Delta E$(HSE).

Figure~\ref{fig:fitU} shows the results of our fitting for
  \LCO\ by using two slightly different approaches. In one case we ran
  the PBEsol+{\sl U} simulations at the HSE-optimized volumes, and in
  the other case we performed volume relaxations at the PBEsol+{\sl U}
  level for each $U$ value considered. It can be seen that such a
  choice does not have a big effect on the computed frequency shifts,
  and both result in the same $U$ value. Hence, the $U$ values reported
  here were obtained by running PBEsol+{\sl U} simulations at the HSE
  volumes.\cite{fn:volume} We obtained $U$= 3.0, 2.8, 2.7, 3.8, and
5.1\,eV, respectively, for \CMO, \SMO, \BMO, \LCO\ and
\LFO.\cite{note1} We then do detailed phonon calculations using this
value of $U$ in the PBEsol+{\sl U} calculations to investigate
spin-phonon coupling effects in \AMO\ and \LMO.

As we will show in Sec.~\ref{sec:results}, we have tested this
proposed scheme and found that it works quite well. In particular,
the phonon frequency shifts $\Delta \omega = \omega_{\rm AFM}-
\omega_{\rm FM}$ computed using PBEsol+{\sl U} with the fitted $U$
are almost the same as those obtained using HSE directly.
Since a previous study has concluded that HSE works ``remarkably well''
for transition-metal oxides,\cite{Cesare} we believe this approach
can be used with confidence.

Finally, we note that the direct HSE calculations can be rather
heavy, even though we only have 10-atom cells; the presence of
magnetic order and the need to calculate the phonons and the
spin-phonon couplings makes the calculations challenging.~\cite{noteTime}
We circumvent this difficulty by splitting a single run
of HSE frozen-phonon calculations into several parallel runs,
each one computing the forces for symmetry-independent atomic
displacements. We then use the forces calculated in these runs
to construct the force-constant matrix, and by diagonalizing
this matrix, we obtain the phonon frequencies and eigenvectors.
Further details of this procedure are presented in the Appendix.

\section{Results and Discussion}
\label{sec:results}

\subsection{Effect of $U$ on the frequency shifts}
\label{sec:U}

Let us begin by showing some representative results of the
$U$-dependence of the phonon frequencies and frequency shifts for the
\AMO\ and \LMO\ compounds. Here we focus on the case of \SMO, which has
been predicted to exhibit a large spin-phonon coupling based on
GGA+$U$ simulations using $U$= 1.7\,eV; more precisely, a giant
frequency shift $\Delta \omega$= 230 \cm1\ has been reported for the
Slater mode at the theoretical equilibrium state.~\cite{Junhee1}

\begin{figure}[t]
    \includegraphics{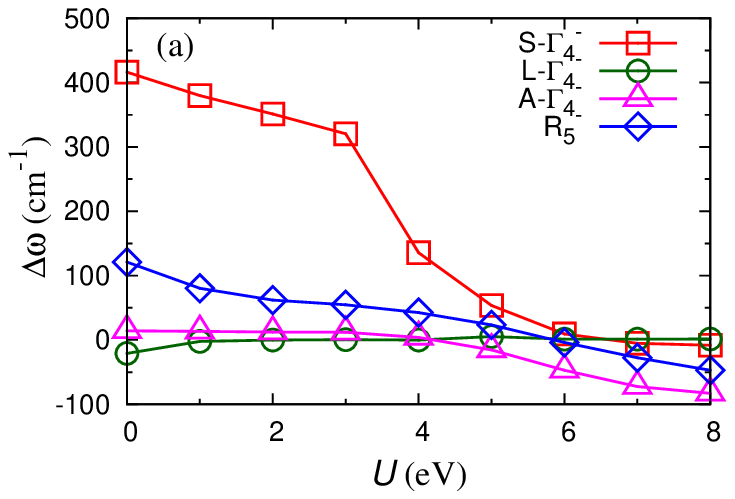}
    \includegraphics{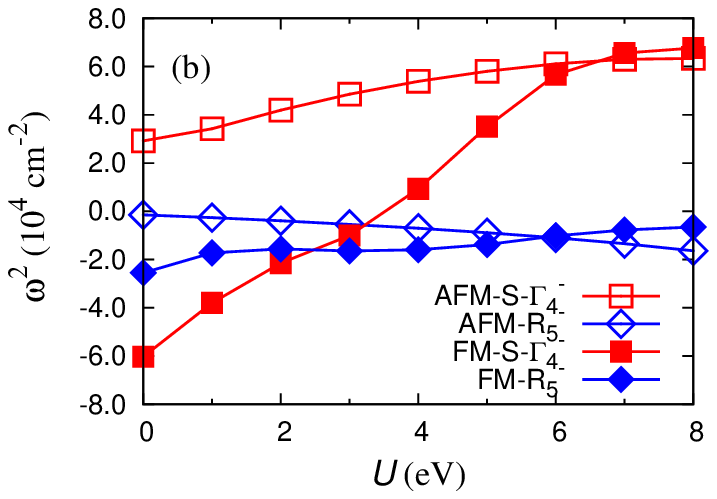}
  \caption{(Color online) Effect of $U$ on spin-phonon coupling
  in \SMO. (a) \G4\ and \R5\ phonon frequency shifts
  ($\Delta \omega = \omega_{\rm AFM}-\omega_{\rm FM}$) versus $U$.
  (b) S-$\Gamma_4^-$ and R$_5^-$ phonon frequencies
  versus $U$ for AFM and FM states.}
  \label{fig:diffU}
\end{figure}

We computed the phonon frequencies and frequency shifts of the ideal
cubic perovskite phase of \SMO\ using $U$ values in the range between
0 and 8\,eV. Figure~\ref{fig:diffU} shows our results for the key
modes that determine the occurrence of ferroelectricity, i.e., the
\G4\ polar phonons and the \R5\ antiferrodistortive (AFD) mode. The
\R5\ mode involves antiphase rotations of the O$_6$ octahedra around
the three principal axes of the perovskite lattice; such a mode is
soft in many cubic perovskites, and it often competes with the FE
instabilities to determine the nature of the low-symmetry phases.

 From Fig.~\ref{fig:diffU}a we see that the \G4\ and \R5\ modes depend
strongly on the chosen $U$. Notably, the frequency shift $\Delta
\omega$ of the S-\G4\ mode can change from 400~\cm1\ to 0~\cm1\ as $U$
increases from 0 to 8\,eV.  The $\Delta \omega$ of the A-\G4\ and
\R5\ modes also depend on the $U$, while the L-\G4\ mode is nearly
insensitive to $U$ within this range of values.  Interestingly, the
large frequency shift of S-\G4\ is related to the strong
$U$-dependence of the FM S-\G4\ mode, as shown in
Fig.~\ref{fig:diffU}b. 

Hence, our results show that the magnitude of the spin-phonon coupling
has a significant dependence on the value of $U$ employed in a DFT+$U$
calculation, and that such a dependence is particularly strong for
some of the key modes determining the structural (FE/AFD)
instabilities of cubic perovskite oxides. It is thus clear that
choosing an appropriate $U$ is of critical importance if we want to
avoid artificially ``strong'' couplings.

\subsection{Spin-phonon coupling in \CMO, \SMO, and \BMO}
\label{subsec:CSB}

\begin{table}[t]
  \caption{ Lattice constant $a$, local Mn magnetic moment $m$, and band
  gap $E_{\rm gap}$ (zero if blank) for \CMO, \SMO\ and \BMO\ from
  PBEsol+$U$ (+$U$) and HSE methods.  }
\begin{ruledtabular}
  \begin{tabular}{ldddd}


&  & \multicolumn{1}{c}{$a$ (\AA)}
   & \multicolumn{1}{c}{$m$ ($\mu_{\rm B}$)}
   & \multicolumn{1}{c}{$E_{\rm gap} (eV) $} \\

    \hline

    \multirow{2}{*}{\CMO-AFM}
    & +U        & 3.73 & 2.79 & 0.35 \\
    & {\rm HSE} & 3.73 & 2.81 & 1.70  \\


    \multirow{2}{*}{\CMO-FM}
    & +U        & 3.74 & 2.84 &  \\
    & {\rm HSE} & 3.73 & 2.81 &   \\

    \cline{1-5}

    \multirow{2}{*}{\SMO-AFM}
    & +U        & 3.80 & 2.80 & 0.40 \\
    & {\rm HSE} & 3.80 & 2.85 & 1.45 \\


    \multirow{2}{*}{\SMO-FM}
    & +U        & 3.81 & 2.89 &  \\
    & {\rm HSE} & 3.81 & 2.91 & 0.00\footnotemark[1] \\

    \cline{1-5}

    \multirow{2}{*}{\BMO-AFM}
    & +U        & 3.91 & 2.86 & 0.13 \\
    & {\rm HSE} & 3.91 & 2.92 & 1.20 \\


    \multirow{2}{*}{\BMO-FM}
    & +U        & 3.93 & 3.14 &  \\
    & {\rm HSE} & 3.93 & 3.13 &  \\

  \end{tabular}
  \end{ruledtabular}
\footnotetext[1]{Half-metallic.}
  \label{tab:Mn-properties}
\end{table}

We studied the cubic phase of \CMO, \SMO, and \BMO\ using the
PBEsol+$U$ approximation with $U$ values for Mn's 3$d$ electrons
(i.e., $U$~=~3.0\,eV for \CMO, $U$~=~2.8\,eV for \SMO, and $U$~=~2.7\,eV
for \BMO) that were determined as described in
Section~\ref{sec:method}. The basic properties that we obtained are
listed in Table~\ref{tab:Mn-properties}. Our PBEsol+{\sl U} method
gives good lattice constants and local Mn magnetic moments compared with
HSE. As regards the
metallic or insulating character, the PBEsol+{\sl U} agrees
qualitatively with the HSE result in most cases, although the band gap
is underestimated. In FM-\SMO\,  there is a clear discrepancy
between PBEsol+$U$ and HSE: the HSE calculations predict
a half-metallic state, while we obtain a metal with PBEsol+{\sl U}.

We have also calculated the Born effective charges for insulating
AFM configurations of \CMO\ (\SMO, \BMO). We obtain $Z_{\rm
Ca/Sr/Ba}$=2.60\,$e$ (2.58\,$e$, 2.73\,$e$),
$Z_{\rm Mn}$=7.35\,$e$ (7.83\,$e$, 9.42\,$e$), $Z_{\rm
O_{\parallel}}$=$-$6.55\,$e$ ($-$6.93\,$e$, $-$8.07\,$e$), $Z_{\rm
O_{\perp}}$=$-$1.70\,$e$ ($-$1.74\,$e$, $-$2.04\,$e$), where $e$ is the elementary
charge and $Z_{\rm O_{\parallel}}$ and $Z_{\rm O_{\perp}}$ refer,
respectively, to the dynamical charges defined for an atomic
displacement parallel and perpendicular to the Mn-O bond.  The
anomalously large $Z_{\rm Mn}$ and $Z_{\rm O_{\parallel}}$
charges of AFM state of \CMO, \SMO\ and \BMO\ are strongly reminiscent of the
results obtained for ferroelectric perovskite oxides,\cite{Ghosez} and
suggest the possible proximity
of a polar instability that might be
triggered by an appropriate external (e.g., strain)
field.\cite{Ghosez2,Junhee1,James}

\begin{table}[t]
  \centering
  \caption{Phonon frequencies and frequency shifts calculated for \CMO,
  in \cm1.  Frequencies calculated using HSE and PBEsol+$U$:
  $\omega^{\rm HSE}$ and $\omega^{+U}$.  Frequency difference between
  HSE and PBEsol+$U$: $\omega^{\Delta}=\omega^{+U} - \omega^{\rm HSE}$.
  Frequency shift: $\Delta \omega = \omega_{\rm AFM}-\omega_{\rm FM}$.
  Difference of frequency shifts between HSE and PBEsol+$U$: $\Delta
  \omega^{\Delta}= \Delta \omega^{+U} - \Delta \omega^{\rm HSE}$.}
  \begin{ruledtabular}
    \begin{tabular}{lrrrrrr}

&  \multicolumn{1}{c}{$\omega_{\rm AFM}^{\rm HSE}$}
&  \multicolumn{1}{c}{$\omega_{\rm AFM}^{\Delta}$}
&  \multicolumn{1}{c}{$\omega_{\rm FM}^{\rm HSE}$}
&  \multicolumn{1}{c}{$\omega_{\rm FM}^{\Delta}$}
&  \multicolumn{1}{c}{$\Delta \omega^{\rm HSE}$}
&  \multicolumn{1}{c}{$\Delta \omega^{\Delta}$}  \\

   \hline

  L-$\Gamma_4^-$ &   126 &--29 &  -81 & --18  & 207 & --10   \\
  S-$\Gamma_4^-$ &   272 &--11 &  186 &--3    & 86  & --8  \\
  $\Gamma_5^-$   &   241 &--48 &  183 & --72  & 58  & 24   \\
  R$_3^-$        &   658 &10   &  636 &  11   & 22  & --1  \\
  R$_4^-$        &   430 &--46 &  413 &--49   & 17  & 2 \\
  R$_5^-$        & --204 &--18 &--218 & --20  & 14  & 2 \\
A-$\Gamma_4^-$   &   569 &35   &  557 & 40    & 12  & --5  \\
  R$_5^+$        &   437 &--11 &  448 & --14  & --11 & 3   \\
   R$_2^-$       &   890 &--27 &  885 & --31  & 5   & 4 \\
  R$_4^-$        &   160 &--20 &  157 & --21  & 3   & 1\\

  \end{tabular}
  \end{ruledtabular}
  \label{tab:CMO-phonon}
\end{table}

The phonons at $\Gamma$ and $R$ for different spin orders for \CMO,
\SMO\ and \BMO\ are shown in Tables~\ref{tab:CMO-phonon},
\ref{tab:SMO-phonon}, and \ref{tab:BMO-phonon}, respectively; the
phonon modes are ordered by descending $\Delta \omega^{\rm HSE}$.  The
first thing to note from the table is that the frequencies obtained
for \SMO's polar modes (i.e., 177, 187 and 494 \cm1) agree well with
available low temperature experiment results (i.e., ~162, 188 and
498~\cm1 taken from Ref.~\onlinecite{Sacchetti}). Thus, our results
provide additional evidence that the HSE scheme renders accurate
phonon frequencies for magnetic oxides. The second thing to note from
these tables is that the phonon-frequency shifts ($\Delta \omega$)
obtained with PBEsol+{\sl U} are in overall good agreement with the
HSE results, which suggests that our strategy to fit the value of $U$
is a good one.

Third, our results offer information about the dependence of the
structural instabilities on the choice of the A-site cation and on the
magnetic arrangement. We find that the AFD instability ($R_5^-$ mode)
becomes weaker and disappears, for both AFM and FM states, as the size
of the A-site cation increases (The effective ionic radii estimated by
Shannon\cite{Shannon} are $r_{\rm Ca} =1.34$\,\AA, $r_{\rm
  Sr}=1.44$\,\AA, and $r_{\rm Ba}=1.61$\,\AA.)  In contrast, the
ferroelectric instability (Slater mode) becomes stronger as the A-site
cation becomes bigger. This is the usual behavior that one would
expect for these two instabilities of the cubic perovskite structure,
and has been recently examined in detail by other
authors.\cite{Tohei,Angel,Ghosez2} It clearly suggests that some of
these compounds could display a magnetically ordered ferroelectric
ground state. In particular, this could be the case for \SMO\ and
\BMO: in these compounds, the AFD instability becomes weaker or even
disappears, so that it can no longer compete with and suppress the FE
soft mode.

\begin{table}[t]
  \centering

  \caption{Calculated phonon frequencies and frequency shifts,
  as in Table \ref{tab:CMO-phonon}, but for \SMO.}

  \begin{ruledtabular}
     \begin{tabular}{lrrrrrr}

&  \multicolumn{1}{c}{$\omega_{\rm AFM}^{\rm HSE}$}
&  \multicolumn{1}{c}{$\omega_{\rm AFM}^{\Delta}$}
&  \multicolumn{1}{c}{$\omega_{\rm FM}^{\rm HSE}$}
&  \multicolumn{1}{c}{$\omega_{\rm FM}^{\Delta}$}
&  \multicolumn{1}{c}{$\Delta \omega^{\rm HSE}$}
&  \multicolumn{1}{c}{$\Delta \omega^{\Delta}$}  \\

   \hline

S-$\Gamma_4^-$ &   177 &    40 &  --96 &--16  & 273 & 56 \\
    R$_5^-$  &     64  & --135 &  --6  &--121 & 70  & --14 \\
$\Gamma_5^-$ &     292 &--38   &  253  &--54  & 39  & 16 \\
    R$_2^-$  &     811 & --26  &  789  &--25  & 22  & --1    \\
    R$_3^-$  &     571 &    12 &  553  &   2  & 18  & 10 \\
    R$_4^-$  &     424 &--47   &  410  &--62  & 14  & 15 \\
L-$\Gamma_4^-$ &   187 & --16  &  177  &--7   & 10  & --9 \\
A-$\Gamma_4^-$ &   494 &   22  &  487  &  17  & 7   & 5 \\
    R$_5^+$  &     412 &--12   &  409  &--12  & 3   & 0 \\
    R$_4^-$  &     159 &--6    &  157  &--7   & 2   & 0 \\

  \end{tabular}
  \end{ruledtabular}
  \label{tab:SMO-phonon}
\end{table}

\begin{figure}[b]
  \begin{center}
    \includegraphics[width= 3.2in] {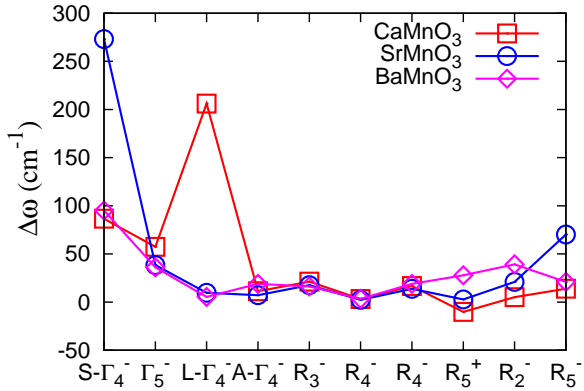}
  \end{center}
  \caption{(Color online) Frequency shifts $\Delta \omega =
      \omega_{\rm AFM}-\omega_{\rm FM}$ in $A$MO$_3$ from HSE
      calculations.}
  \label{fig:shift_AMO}
\end{figure}

The phonon frequency shift $\Delta \omega$ between the AFM and FM
magnetic orders is shown in Fig.~\ref{fig:shift_AMO}. We find that the
Slater modes exhibit a very considerable spin-phonon coupling for all
three compounds, with $\Delta\omega\gtrsim$~100~cm$^{-1}$. A sizable
effect is also obtained for the $\Gamma_5^-$ mode
($\Delta\omega\sim$~50~cm$^{-1}$) in all cases. Additionally, the Last
mode shows a very large effect for \CMO, and the spin-phonon
coupling for \R5\ is significant in the case of \SMO. The coupling is
relatively small, or even negligible, for all other modes.

\begin{table}[t]
  \centering

  \caption{Calculated phonon frequencies and frequency shifts,
  as in Table \ref{tab:CMO-phonon}, but for \BMO.}

  \begin{ruledtabular}
     \begin{tabular}{lrrrrrr}

&  \multicolumn{1}{c}{$\omega_{\rm AFM}^{\rm HSE}$}
&  \multicolumn{1}{c}{$\omega_{\rm AFM}^{\Delta}$}
&  \multicolumn{1}{c}{$\omega_{\rm FM}^{\rm HSE}$}
&  \multicolumn{1}{c}{$\omega_{\rm FM}^{\Delta}$}
&  \multicolumn{1}{c}{$\Delta \omega^{\rm HSE}$}
&  \multicolumn{1}{c}{$\Delta \omega^{\Delta}$}  \\

   \hline

S-$\Gamma_4^-$   &--274&  117  & --369 &  227  & 95 & --110  \\
 R$_2^-$         &670  &--28   &  631  &--59   & 39 & 31  \\
$\Gamma_5^-$     &317  &--34   &  281  &--28   & 36 & --6 \\
R$_5^+$          &363  &--13   &  335  &--22   & 28 & 9 \\
R$_5^-$          &221  &--28   &  200  &--67   & 21 & 39 \\
R$_4^-$          &403  &--48   &  384  &--60   & 19 & 12\\
A-$\Gamma_4^-$   &423  &--12   &  404  &  11   & 19 & --22 \\
R$_3^-$          &423  &  16   &  407  &--20   & 16 & 36 \\
L-$\Gamma_4^-$   &200  &--6    &  195  &--6    & 5  & 0 \\
R$_4^-$          &150  &--5    &  147  &--6    & 3  & 1\\

  \end{tabular}
  \end{ruledtabular}
  \label{tab:BMO-phonon}
\end{table}

\begin{figure}[b]
  \begin{center}
   \includegraphics[width= 2.6in] {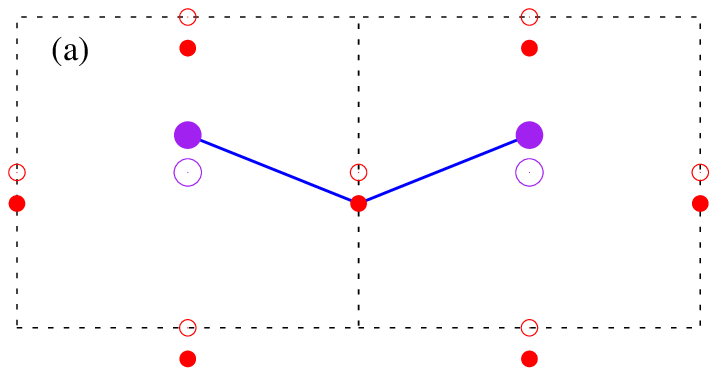}
    \includegraphics[width= 2.6in] {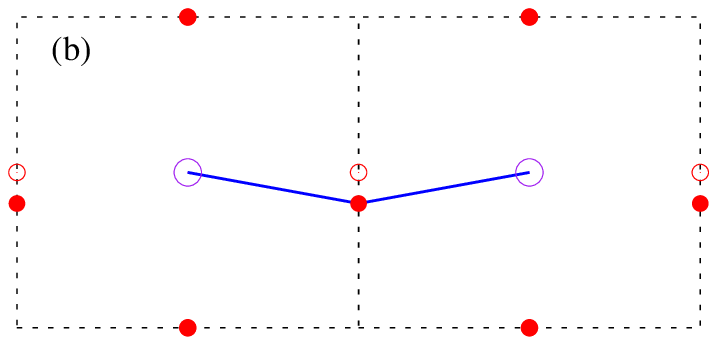}

  \end{center}
  \caption{(Color online) Changes to metal--O--metal bond angle
  (solid blue line) resulting from Slater (a) and $\Gamma_5^-$ (b)
  phonon modes. Open and solid dots indicate ideal and displaced
  positions respectively. Small red dots are oxygen; larger
  purple dots are metal atoms.}
  \label{fig:MOM}
\end{figure}

\begin{table*}[t]
  \caption{Lattice constant $a$, local magnetic moment $m$, band
  gap $E_{\rm gap}$, dielectric constant $\varepsilon$, and
  Born effective charge $Z$ for \LCO, \LFO\ and \LCFO\ (LCFO)
  from GGA+$U$ (+$U$) and HSE methods.  }

  \begin{ruledtabular}
  \begin{tabular}{lcccccccccc}

    &  & $a$ (\AA) & $m$ ($\mu_{\rm B}$) & $E_{\rm gap}$ (eV) &
    $\varepsilon_{\infty}$ & $\varepsilon_0$ & $Z_{\rm La}$ ($e$) &
    $Z_{\rm Cr/Fe}$ ($e$) & $Z_{\rm  O_{\parallel}}$ ($e$)& $Z_{\rm
    O_{\perp}}$ ($e$) \\
    \hline

    \multirow{2}{*}{\LCO-AFM}
    & +$U$& 3.87 & 2.86 & 2.01 & 6.12 & 93.71 &4.53 & 3.34 & $-$3.50 & $-$2.19\\
    & HSE & 3.87 & 2.75 & 3.06 &  &    &   &    &     &  \\
    \multirow{2}{*}{\LCO-FM}
    & +$U$& 3.88 & 2.84 & 1.70 & 6.08 & 236.19 &4.52 & 3.61 & $-$3.62 & $-$2.26\\
    & HSE & 3.88 & 2.74 & 2.49 &    &      &   &    &     &  \\

    \cline{1-11}

    \multirow{2}{*}{\LFO-AFM}
    & +$U$& 3.90  & 4.19 & 2.20 & 6.37 &  & 4.47 & 3.75 & $-$3.41 & $-$2.43\\
    & HSE & 3.91 & 4.10 & 3.25 &    &    &    &    &     &  \\
    \multirow{2}{*}{\LFO-FM}
    & +$U$& 3.91 & 4.38 & 1.57 & 6.19 &  &4.48 & 4.02 & $-$3.59 & $-$2.46\\
    & HSE & 3.93 & 4.26 & 2.20 &    &    &   &    &     &  \\

    \cline{1-11}

    \multirow{2}{*}{LCFO--AFM}
    & +$U$& 3.88 & 2.71/4.30 & 1.98 & 6.35 & 713.57  & 4.51  & 3.02/4.22 & $-$3.50 & $-$2.33  \\
    & HSE & 3.88 & 2.63/4.19 & 2.78 \\
    \multirow{2}{*}{LCFO--FM}
    & +$U$  & 3.89 & 3.01/4.28 & 2.01 & 6.18 & 262.40 & 4.49 & 3.21/3.87 & $-$3.45 & $-$2.30 \\
    & HSE   & 3.89 & 2.89/4.18 & 3.04 \\

  \end{tabular}
  \end{ruledtabular}

  \label{tab:La-properties}
\end{table*}

We will not attempt any detailed interpretation of our quantitative
results here, but some words are in order. In the recent literature,
it has often been claimed that AFD distortions are expected to couple
strongly with the magnetic structure of perovskite oxides, as they
control the metal--oxygen--metal angles that are critical to determine
the magnitude of the magnetic interactions that are dominant in
insulators.\cite{Ref33} [The hopping parameters between oxygen (2$p$)
  and transition metal (3$d$) orbitals are strongly dependent on such
  angles, as are the effective hoppings between 3$d$ orbitals of
  neighboring transition metals. As we know form the
  Goodenough-Kanamori-Anderson rules, the super-exchange interactions
  are strongly dependent on such hoppings.] In principle, such an
effect should be reflected in our computed frequency shifts; for such
\R5\ modes we obtain $\Delta\omega$ values ranging between
10~cm$^{-1}$ and 70~cm$^{-1}$. Interestingly, following the same
argument, one would conclude that the Slater and $\Gamma_5^-$ modes
should have a similar impact on the magnetic couplings in these
materials, as both involve changes in the metal--oxygen--metal angles
due to the relative displacement of metal and oxygen atoms, as shown
in Fig.~\ref{fig:MOM}. Further, the Slater modes also affect the
metal--oxygen distances, as they involve a significant shortening of
some metal--oxygen bonds (Fig.~\ref{fig:MOM}a). Hence, from this
perspective, the large $\Delta\omega$ values obtained in our
calculations for these two types of modes are hardly
surprising. Finally, the result obtained for the Last mode in the case
of \CMO\ (i.e., $\Delta\omega\approx$~200~cm$^{-1}$) is clearly
anomalous and unexpected according to the above arguments. We
speculate that it may be related to a significant Ca--O interaction
that alters the usual super-exchange mechanism and favors a FM
interaction. While unusual, effects that seem similar have been
reported previously for other compounds.\cite{iniguez05,kim10}

\subsection{Spin-phonon coupling in \LCO, \LFO, and \LCFO }
\label{sect:double}

\begin{table}[b]
  \centering

  \caption{Calculated phonon frequencies and frequency shifts,
  as in Table \ref{tab:CMO-phonon}, but for \LCO.}

  \begin{ruledtabular}
     \begin{tabular}{lrrrrrr}

&  \multicolumn{1}{c}{$\omega_{\rm AFM}^{\rm HSE}$}
&  \multicolumn{1}{c}{$\omega_{\rm AFM}^{\Delta}$}
&  \multicolumn{1}{c}{$\omega_{\rm FM}^{\rm HSE}$}
&  \multicolumn{1}{c}{$\omega_{\rm FM}^{\Delta}$}
&  \multicolumn{1}{c}{$\Delta \omega^{\rm HSE}$}
&  \multicolumn{1}{c}{$\Delta \omega^{\Delta}$}  \\

   \hline

$\Gamma_5^-$    & 230  &  --24 &  178 & --35 & 52 &  11  \\
S-$\Gamma_4^-$  & 361  &  --26 &  317 & --27 & 44 &  1   \\
L-$\Gamma_4^-$  &  69  &    0  &  48  & --3  & 21 &  3   \\
R$_4^-$         & 389  &  --29 &  372 & --30 & 17 &  1   \\
R$_5^+$         & 414  &  2    & 430  &   0  &--16&  2  \\
A-$\Gamma_4^-$  & 659  &  32   & 644  &   30 & 15 &  2  \\
R$_5^-$         & --213&  1    &--228 &  1   & 15 &  0  \\
R$_3^-$         & 683  &  5    & 669  &  5   &  14&  0  \\
R$_4^-$         & 89   &  --4  &  81  &  --4 &  8 &  0   \\
 R$_2^-$        & 840  &  --17 & 843  & --19 &--3 &  2  \\

  \end{tabular}
  \end{ruledtabular}
  \label{tab:LCO-phonon}
\end{table}

In the previous Section we have shown how spin-phonon coupling effects
can crucially depend on the nature of the {\sl A}-site cations
in {\sl A}MnO$_3$ compounds.  Now we will focus on the change of
{\sl B}-site cation to investigate the spin-phonon couplings in
the La$M$O$_3$ compounds ($M$=Cr, Fe) and the double perovskite \LCFO. For our
PBEsol+$U$ calculations we used $U$~=~3.8\,eV for Cr and $U$~=~5.1\,eV
for Fe, which were determined as described in
Section~\ref{sec:method}. We obtained these $U$'s from calculations
for \LCO\ and \LFO\, and used the same values for the PBEsol+$U$ study
of \LCFO.

\begin{table}[b]
  \centering

  \caption{Calculated phonon frequencies and frequency shifts,
  as in Table \ref{tab:CMO-phonon}, but for \LFO.}

  \begin{ruledtabular}
     \begin{tabular}{lrrrrrr}

&  \multicolumn{1}{c}{$\omega_{\rm AFM}^{\rm HSE}$}
&  \multicolumn{1}{c}{$\omega_{\rm AFM}^{\Delta}$}
&  \multicolumn{1}{c}{$\omega_{\rm FM}^{\rm HSE}$}
&  \multicolumn{1}{c}{$\omega_{\rm FM}^{\Delta}$}
&  \multicolumn{1}{c}{$\Delta \omega^{\rm HSE}$}
&  \multicolumn{1}{c}{$\Delta \omega^{\Delta}$}  \\

   \hline

   R$_2^-$     & 796  &--38  &  822 &  --40 &-26 &  2 \\
A-$\Gamma_4^-$ & 645  &--7   & 630  &  --4  &15  &  --3 \\
S-$\Gamma_4^-$ & 259  &  4   & 247  &   4   &12  &  0 \\
     R$_3^-$   & 549  & --11 &  536 &  --8  & 13 &  --3 \\
     R$_4^-$   & 390  &--24  &  379 &  --23 & 11 & --1\\
L-$\Gamma_4^-$ &--81  &  12  & --92 &   9  & 11 &  3\\
     R$_5^-$   &--237 &   9  & --247&  7    &10  &  2\\
     R$_4^-$   &  66  &--2   &   56 &  --2  & 10 &  0 \\
     R$_5^+$   & 346  &--5   &  355 &  --6  &--9 &  1 \\
$\Gamma_5^-$   & 120  &--3   &  115 &  --10  & 5  &  7\\

  \end{tabular}
  \end{ruledtabular}
  \label{tab:LFO-phonon}
\end{table}

The basic computed properties of \LCO, \LFO\ and \LCFO\ are presented
in Table~\ref{tab:La-properties}.  In all cases, we obtain insulating
solutions for AFM and FM spin orders, both from HSE and PBEsol+$U$
calculations. Therefore, we also calculated the Born effective charges
and optical dielectric constants within PBEsol+$U$ by using
density-functional perturbation theory as implemented in VASP. The
static dielectric constants were also calculated for compounds which
do not have an unstable polar mode. It is evident that $Z_{\rm La}$
and $Z_{\rm O_{\perp}}$ are essentially identical for the three
compounds and insensitive to the spin order. On the other hand, the
$Z_{\rm Cr/Fe}$ and $Z_{\rm O_{\parallel}}$ charges of \LCO\ and
\LFO\ increase in magnitude when the spin arrangement changes from AFM
to FM, and decrease in the case of the double perovskite \LCFO.
Table~\ref{tab:La-properties} also shows that the optical dielectric
constants $\varepsilon_{\infty}$ are very close to 6 for these three
materials, and are independent of the magnetic order. However, the
static dielectric constants $\varepsilon_0$ of \LCO\ and \LCFO\ change
very significantly when moving from AFM to FM. Also, the static
dielectric constant of AFM-\LCFO\ is very
large due to the very small frequency of the Last phonon mode. The
static dielectric constants of \LFO\ are not shown in
Table~\ref{tab:La-properties}
because the Last modes are unstable, and strictly speaking they are
not well defined. (Roughly speaking, in all such cases we would have
$\varepsilon_{0}\rightarrow\infty$, as the cubic phase is unstable with
respect to a polar distortion.)

In Tables~\ref{tab:LCO-phonon} and \ref{tab:LFO-phonon} we show the
phonon frequencies as calculated at the HSE and PBEsol+{\sl U} levels
for \LCO\ and \LFO\ respectively. Further, Table~ \ref{tab:LCFO-phonon}
shows the PBEsol+{\sl U} results for \LCFO. As was the case
for the {\sl A}MnO$_3$
compounds of the previous section, we find here as well that the
AFM-FM frequency shifts computed with our PBEsol+{\sl U} scheme
reproduce well the HSE results.

Our results also show that the phonons of the \LMO\ compounds exhibit
some features that differ from those of the \AMO\ materials. First,
the octahedral rotation mode (\R5) is unstable for all the La-based
compounds, and it is largely insensitive to the nature of the {\sl
  B}-site cation. Second, the energetics of the FE modes is very
different. In the case of the Mn-based compounds, the Slater mode is
the lowest-frequency mode, and in some cases it becomes unstable, thus
inducing a polarization, just as the unstable Slater mode induces
ferroelectricity in BaTiO$_3$.\cite{Hlinka} However, in the La compounds
the lowest-frequency mode is the Last phonon mode, and in the
cases of \LFO\ and \LCFO\ this Last mode is unstable and might induce
ferroelectricity; this situation is more analogous to what occurs in
PbTiO$_3$~\cite{Hlinka} or BiFeO$_3$.

\begin{table}[b]
  \centering

  \caption{Calculated phonon frequencies and frequency shifts,
  as in Table \ref{tab:CMO-phonon}, but for \LCFO\ from
  PBEsol+$U$.~\cite{noteMode} }

  \begin{ruledtabular}
     \begin{tabular}{lrrr}
 & $\omega_{\rm AFM}$ & $\omega_{\rm FM}$ & $\Delta \omega$  \\
   \hline

 L-$\Gamma_4^-$ &     31   &    45    &   --14  \\
 $\Gamma_5^-$   &    161   &    178   &   --17  \\
     R$_5^+$    &    383   &    400   &   --17  \\
S-$\Gamma_4^-$  &    284   &    297   &   --13  \\
     R$_3^-$    &    629   &    616   &    13  \\
    R$_2^-$     &    803   &    797   &    6  \\
    R$_5^-$     &  --225   &  --222   & --3  \\
   R$_4^-$      &    362   &    360   &    2  \\
A-$\Gamma_4^-$  &    674   &    676   &    --2 \\
    R$_4^-$     &     79   &     79   &    0  \\

  \end{tabular}
  \end{ruledtabular}
  \label{tab:LCFO-phonon}
\end{table}

\begin{figure}[t]
  \begin{center}
    \includegraphics[width= 3.2in] {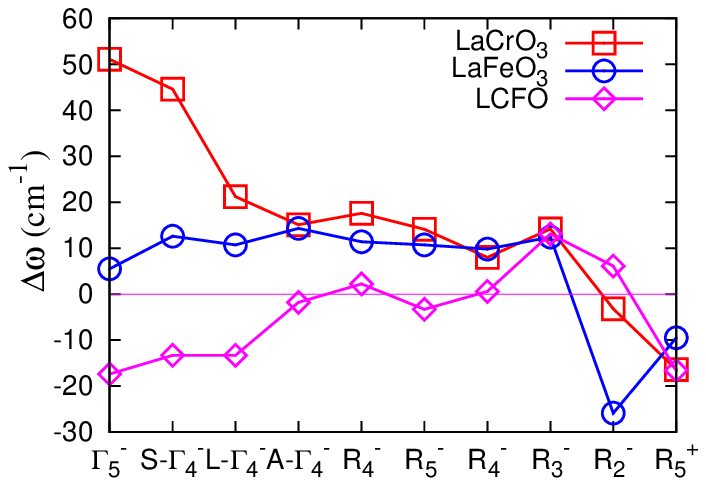}
  \end{center}
  \caption{(Color online) Frequency shifts  for La$M$O$_3$ (HSE
  for \LCO\ and \LFO; PBEsol+$U$ for \LCFO).}
  \label{fig:shift_LMO}
\end{figure}

The spin-phonon coupling effects computed for the \LMO\ compounds are
given in Tables~\ref{tab:LCO-phonon}, \ref{tab:LFO-phonon}, and
\ref{tab:LCFO-phonon}, and are summarized in
Fig.~\ref{fig:shift_LMO}. Here, the first thing to note is that the
magnitude of the effects is significantly smaller than for the
\LMO\ compounds (note the different scales of
Figs.~\ref{fig:shift_AMO} and \ref{fig:shift_LMO}).  Second, we
observe that the largest effects are associated with the $\Gamma_5^-$
and Slater modes (with $\Delta\omega\approx$~45~cm$^{-1}$ for
\LCO). This is consistent with the point made above that these modes
disrupt the metal--oxygen--metal super-exchange paths. In comparison,
in this case we obtain a relatively small effect for the \R5\ modes,
which show $\Delta\omega$ values ($\lesssim$~20~cm$^{-1}$) that are
comparable to those computed for most of the phonon modes
considered. Finally, for \LCO\ and \LFO\ we observe that the $\Gamma$
phonon frequencies decrease as the spin order changes from AFM to FM,
in line with what was observed for the \AMO\ compounds. In contrast,
the $\Gamma$ modes of double-perovskite \LCFO\ increase in frequency
when the spin order changes to FM.

\subsection{Prospects for the design of new multiferroics}
\label{sect:multiferroics}

Let us now discuss several possible implications of our results
regarding the design of novel multiferroic materials.

\subsubsection{\BMO-based materials}

As already mentioned in the Introduction, the spin-phonon coupling can
trigger multiferroicity in some materials if, by application of an
epitaxial strain or other perturbation, a FM-FE state can be
stabilized with respect to the AFM-PE ground state.\cite{Craig1}
Indeed, the three investigated {\sl A}MnO$_3$ compounds have been
suggested as candidates to exhibit strain-induced multiferroicity by
three research groups.~\cite{Ghosez2,Junhee1,James} According to our
calculations, we can tentatively suggest that \BMO\ might show
multiferroicity even without strain applied, provided the material can
be grown as a distorted perovskite (the most stable polymorph of
\BMO\ adopts instead a structure with face-sharing octahedra
\cite{Chamberland,James}). Note that it may
be possible to enhance the stability of \BMO's cubic perovskite
phase by partial substitution of Ba by Ca or Sr. In fact, ideally one
would try to obtain samples of Ba$_{1-x}$Sr$_x$MnO$_3$ or
Ba$_{1-x}$Ca$_x$MnO$_3$ with $x$ large enough to stabilize the
perovskite phase, and small enough for the FE instability associated
to the FM Slater mode to dominate over the \R5\ and AFM Slater
modes. Apparently this strategy to obtain new multiferroics has
recently been realized experimentally by Sakai {\sl et
  al}.\cite{Sakai} in Ba$_{1-x}$Sr$_x$MnO$_3$ solid solutions. An
alternative would be to consider \CMO/\BMO\ or
\SMO/\BMO\ superlattices with varying ratios of the pure compounds,
and perhaps tuning the misfit strain via the choice of substrate.

Another intriguing possibility pertains to the magnetic response of
the {\sl A}MnO$_3$ compounds that display an AFM-PE ground state and a
dominant polar instability of their FM phase. Again, this could be the
case for some Ba$_{1-x}$Sr$_x$MnO$_3$ and Ba$_{1-x}$Ca$_x$MnO$_3$ solid
solutions with an appropriate choice of $x$. By applying a magnetic
field to such compounds, it might be possible to switch them from the AFM
ground state to a FM spin configuration and, as a result, induce a
ferroelectric polarization.

\subsubsection{Double-perovskite \LCFO}

The double perovskite \LCFO\ has been intensively studied to examine
its possible magnetic order through $3d^3-3d^5$
superexchange. However, its magnetic ground state has long been
debated. Pickett \textit{et al.}~\cite{Pickett} predicted that the
ferrimagnetic (FiM) ground state with a net spin moment of
2\,$\mu_{\rm B}$/f.u.\ is more stable than the FM one with
7\,$\mu_{\rm B}$/f.u. (This FiM order can be viewed as a G-AFM
configuration in which, for example, all Fe spins are pointing up and
all Cr spins are pointing down.) However, from GGA and LDA+$U$
calculations, Miura \textit{et al.}\ found that the ground-state
magnetic ordering of \LCFO\ is FiM in GGA, but that even a small $U$
in LDA+$U$ makes it FM.~\cite{Miura} The experimental picture is also
unclear.  Ueda \textit{et al.}\ have grown a (111)-oriented
\LCO/\LFO\ superlattice which exhibits FM ordering, although the
measured saturation magnetization is much smaller than
expected. \cite{Ueda} Very recently, Chakraverty \textit{et
  al.}\ reported epitaxial \LCFO\ double-perovskite films
grown by pulsed-laser deposition, and their sample
exhibits FiM with a saturation magnetization of $2.0\pm
0.15\,\mu_{\rm B}$/f.u.\ at 5\,K.~\cite{Chakraverty}

Our HSE calculations for \LCFO\ with the atomic (oxygen) positions
relaxed in the cubic structure
shows the magnetic ground state has a FiM spin
pattern leading to a FiM structure with a net magnetization of
1.56\,$\mu_{\rm B}$/f.u. This is consistent with
LDA~\cite{Pickett} and GGA~\cite{Miura} calculations, as well as
being heuristically consistent with the experimental report of
AFM ordering in \LCFO\ solid-solution films.\cite{Chakraverty}
However, our HSE
calculation shows that the energy difference between the FiM and FM
states is very small: FiM is only 0.8\,meV/f.u.\ lower in energy than
FM.  In addition, the GGA+$U$ with $U$ fitted to \LCO\ and
\LFO\ ($U$=3.8 and 5.1\,eV for Cr and Fe, respectively) results
in a FM magnetic ground state having a total energy
34.9\,meV lower than that of the FiM.
By doing a fitting of the $U$ parameters for Cr and Fe directly
to $E_{\rm AFM}-E_{\rm FM}$ of \LCFO\ as computed by PBEsol+$U$
and HSE, instead of for \LCO\ and \LFO\ separately, we find
parameters of $U$=3.0 and 4.1\,eV for Cr and Fe respectively (the
phonon properties predicted from these are very close to
our previous results).  Using these parameters, we find that the
FiM ground state is 0.7\,meV lower in energy than the FM state.
For comparison, a straight PBEsol calculation (with $U=0$)
predicts that the energy of the FiM ground state is 596\,meV/f.u.\ lower
than that of the FM state. Clearly, $U$ should
be chosen carefully in order to obtain the correct ground state
of \LCFO.

According to our HSE calculations, the magnetic ground state
of \LCFO\ is FiM, but with the FM state lying only very
slightly higher in energy.  This may be the reason why questions
about the magnetic ground state of \LCFO\ have long been debated; the
energy difference is so small that external perturbations (e.g.,
the epitaxial strain in a superlattice~\cite{Ueda}) or variations in
$U$ between different LDA+$U$~\cite{Miura} and GGA+$U$
calculations may bring the FM energy below that of the AFM.
Taken together with our results, shown in Table~\ref{tab:LCFO-phonon},
that the Last mode is close to going soft in this material,
this suggests that \LCFO\ might be a good candidate for
a material in which multiferroic phase transitions could be induced,
similar to what was
shown for \SMO~\cite{Junhee1} and SrCoO$_3$.~\cite{strain4}
Because the spin ordering is so delicate, it seems likely that
a small misfit strain could be enough to trigger such a transition.

\section{Summary and conclusions}
\label{sec:summ}

In this work, we have studied the spin-phonon coupling for
transition-metal oxides within density-functional theory.  From the
computational point of view, an accurate description of the
electronic, structural, and vibrational properties on an equal footing
is a prerequisite for a reliable study of the coupling between spins
and phonon.  Taking note of the increasing evidence that hybrid
functionals are suitable for this task, we have adopted the
Heyd-Scuseria-Ernzerhof (HSE) screened hybrid functional for the
present work.  However, the accuracy of the HSE results comes at the
cost of an increase of computational load, so that a full
frozen-phonon calculation of the phonon modes remains prohibitively
expensive in many cases.  We propose to overcome this limitation by
carrying out calculations at the DFT+$U$ level using $U$ parameters
that have been fitted to HSE results for total-energy differences
between spin configurations. Our results show that the resulting
DFT+$U$ scheme reproduces the HSE results very accurately, especially
in regard to the spin-phonon couplings of interest here.

As regards the direct HSE phonon calculations, we have developed an
approach in which we split the calculation into separate, simultaneous
frozen-phonon calculations for different symmetry-adapted displacement
patterns, and then combine the results to calculate and diagonalize
the dynamical matrix, thus accelerating these calculations
significantly.

Our important results can be summarized as follows.  First, we have
shown that the choice of $U$ is a big concern in such studies, since
the spin-phonon coupling can depend very strongly on $U$.  As an
alternative to extracting $U$ from experimental studies, we propose
here to obtain it by fitting to HSE calculations as illustrated above.
Second, we have studied \CMO, \SMO\ and \BMO, focusing on trends in
the spin-phonon coupling due to the increase of the $A$ cation size.
Based on the strain couplings and the spin-phonon interactions, we
suggest theoretically that \BMO\ is more likely to show
ferroelectricity under tensile strain and, furthermore, that A-site
substitution by a cation with smaller size may induce multiferroicity
even without external strain.  Third, we find that in the $A$MnO$_3$
materials class with $A$=Ca, Sr, and Ba, the frequency shift decreases
as the A cation radius increases for the $\Gamma$ phonons, while it
increases for R-point phonons.  Fourth, we have shown that changing
B-site cations may also have important effects on the dielectric
properties: in La$M$O$_3$ with $M$=Cr, Fe, and Cr/Fe, the phonon
frequencies at $\Gamma$ decrease as the spin order changes from AFM to
FM for LaCrO$_3$ and LaFeO$_3$, but they increase for the double
perovskite La$_2$(CrFe)O$_6$. Finally, we have shown that the polar
phonon modes of the investigated perovskites tend to display the
largest spin-phonon couplings, while modes involving rotations of the
O$_6$ octahedra present considerable, but generally smaller,
effects. Such observations may be relevant as regards current efforts
to obtain large magnetostructural (and magnetoelectric) effects.

We hope that our study will stimulate further work leading to
rational design and strain engineering of multiferroicity using
spin-phonon couplings.

\appendix*

\section{Efficient phonon calculations with hybrid functionals}

Even though we only have a 10-atom cell, we found that it can be quite
expensive to use the HSE functional to carry out the needed
spin-polarized calculations of phonon properties.~\cite{noteTime} 
We overcome this limitation as follows.  First, we use symmetry
to limit ourselves to sets of displacements that will block-diagonalize
the force-constant matrix.  For example, for the polar modes
we move the cations along $x$ and the O atoms along $x$ and $y$.
We then carry out self-consistent calculations on these displaced
geometries, and from the forces we construct the relevant block
of the force-constant matrix.  Second, while the standard VASP
implementation uses a ``central difference'' method in which
ions are displaced by small amounts in both positive and negative
displacements, we save some further effort by displacing only in the
positive direction.  Finally, we note that the forces resulting
from each pattern of atomic displacements can be calculated
independently, allowing us to split the calculation in parallel
across independent groups of processors and thus further reduce
the wall-clock time.

We have checked the accuracy of this approach for FM-\LCO\
using PBEsol+$U$ ($U$=3.8\,eV) and for G-AFM-\BMO\ using HSE. For
each case, we compared the results of the standard implementation
of the VASP frozen-ion calculation of phonon frequencies with
the revised approach described above.  We take the ion
displacements to be 0.015\,\AA\ in all cases.
We find that the RMS error of ten
different phonon frequencies is 1.2\,\cm1 for
PBEsol+$U$ and 7.3\,\cm1 for HSE.  These results suggest that this
method has acceptable accuracy with reduced computational
cost. We propose that it could be used also for cases of lower
symmetry and larger cells, thus making the HSE phonon calculations
at $\Gamma$ affordable in general.
  
Recently, a revised Perdew--Burke--Ernzerhof functional base on
PBEsol, which we refer to as HSEsol, was designed to yield more accurate
equilibrium properties for solids and their surfaces. Compared to
HSE, significant improvements were found for lattice constants and
atomization energies of solids.~\cite{HSEsol}  We also checked
the effect of using HSEsol on our calculations, as shown in
Table~\ref{tab:HSEsol}.  From this table, we can see that the
phonons calculated with HSEsol are very close to those from HSE.  

\begin{table}[t]
  \centering
  \caption{Comparison of phonon frequencies (\cm1) of \SMO\ calculated
  from HSE and HSEsol (sol).}
  \begin{ruledtabular}
\begin{tabular}{lrrrrrrrrr}

  & \multicolumn{1}{c}{$\omega_{\rm AFM}^{\rm HSE}$}
 & \multicolumn{1}{c}{ $\omega_{\rm AFM}^{\rm sol} $ }
& \multicolumn{1}{c}{ $\omega_{\rm FM}^{\rm HSE} $}
& \multicolumn{1}{c}{ $\omega_{\rm FM}^{\rm sol} $}
& \multicolumn{1}{c}{ $\Delta \omega^{\rm HSE} $}
& \multicolumn{1}{c}{ $\Delta \omega^{\rm sol} $} \\

   \hline

S-$\Gamma_4^-$ & 177  &  166 & --96 &--129 & 273 & 295  \\
$\Gamma_5^-$ &   292  &  290 &  253 &  250 & 38  &  40  \\
L-$\Gamma_4^-$ & 187  &  179 &  177 &  172 & 10  &  7  \\
A-$\Gamma_4^-$ & 494  &  485 &  487 &  477 & 7   &  8   \\
    R$_4^-$  &   159  &  156 &  157 &  154 & 2   &  3   \\
    R$_5^+$  &   412  &  408 &  409 &  405 & 3   &  3   \\
    R$_4^-$  &   424  &  419 &  410 &  406 & 14  &  14 \\
    R$_3^-$  &   571  &  558 &  553 &  539 & 18  &  19  \\
    R$_5^-$  &   64   &  71  & --6  &  26  & 70  &  45  \\
    R$_2^-$  &   811  &  803 &  789 &  782 & 21  &  20 \\

  \end{tabular}
  \end{ruledtabular}
  \label{tab:HSEsol}
\end{table}

\acknowledgments

This work was supported by ONR Grant 00014-05-0054, by Grant Agreement
No.~203523-BISMUTH of the EU-FP7 European Research Council, and by
MICINN-Spain (Grants No.~MAT2010-18113, No.~MAT2010-10093-E, and
No.~CSD2007-00041). Computations were carried out at the Center for
Piezoelectrics by Design.  J.H.\ acknowledges travel support from
AQUIFER Programs funded by the International Center for Materials
Research at UC Santa Barbara. We thank Jun Hee Lee, Claude Ederer
and Karin Rabe for useful discussions.



\begin{thebibliography}{}
\bibitem{fn:volume} In the materials considered here, the
  equilibrium volumes obtained from HSE and PBEsol+{\sl U} are
  essentially the same (see Tables~\protect\ref{tab:Mn-properties}
  and \protect\ref{tab:La-properties}); hence, the determined {\sl
    U}'s are essentially independent from our imposing the HSE-volume
  in the PBEsol+{\sl U} calculations or not. However, in the general
  case this may not be true, and it seems reasonable to perform
  structural relaxations at the DFT+{\sl U} level for each {\sl U}
  considered.
\bibitem{note1} We found two $U$ values, $U$ = 2.2 and 5.1\,eV,
  that generate the same $\Delta E^{+U}= \Delta E^{\rm HSE}$ for
  \LFO. We chose 5.1\,eV because the band gap and magnetic moment
  are in better agreement with HSE calculations. We also found
  the FM energy to cross below the AFM one for $U<0.8$\,eV.
\bibitem{noteTime} In our case it takes about 6 days to obtain
  the phonon frequency for G-AFM \BMO\ using a direct HSE
  calculation with 96 cores (AMD
  Opteron 6136 2.4GHz with Infiniband network), but less
  then 0.5 hours using PBEsol+U with only 16 cores.
\bibitem{noteMode} Because the symmetry is lower than for a $Pm\bar 3 m$
  perovskite, some modes of the double perovskite \LCFO\ listed in
  Table~\ref{tab:LCFO-phonon} are combinations of two modes. We
  list the dominanant contribution in the Table.  For
  example, the mode with frequency 383\,\cm1\ in the AFM structure
  is 86.5\% R$_5^+$ and 13.5\% $\Gamma_4^-$, so we denote it as R$_5^+$.
\end{thebibliography}
\end{document}